# Variance component score test for multivariate change point detection with applications to mobile health

Melissa Lynne Martin[1*], Juliette Brook[2-4], Sage Rush[2-4], Theodore D. Satterthwaite[2-4], Ian J. Barnett[1]

1. Department of Biostatistics, Epidemiology, and Informatics, University of Pennsylvania
2. Department of Psychiatry, University of Pennsylvania
3. Penn Lifespan Informatics and Neuroimaging Center (PennLINC), University of Pennsylvania
4. Penn-Children's Hospital of Philadelphia Lifespan Brain Institute, University of Pennsylvania

*Corresponding Author

martin30@pennmedicine.upenn.edu

## Abstract

Multivariate change point detection is the process of identifying distributional shifts in time-ordered data across multiple features. This task is particularly challenging when the number of features is large relative to the number of observations. This problem is often present in mobile health, where behavioral changes in at-risk patients must be detected in real time in order to prompt timely interventions. We propose a variance component score test (VC*) for detecting changes in feature means and/or variances using only pre-change point data to estimate distributional parameters. Through simulation studies, we show that VC* has higher power than existing methods. Moreover, we demonstrate that reducing bias by using only pre-change point days to estimate parameters outweighs the increased estimator variances in most scenarios. Lastly, we apply VC* and competing methods to passively collected smartphone data in adolescents and young adults with affective instability.

## 1 Introduction

Mobile health (mHealth) is a broad term that generally refers to the use of mobile devices, such as smartphones and wearable devices, to aid in the monitoring and treatment of various health conditions. mHealth has been particularly useful in the area of mental health, as it provides insights into behavioral patterns through low-burden smartphone data collection.[1–4] Many mental health disorders are associated with adverse events, such as suicidal ideation in patients with major depressive disorder and hospitalization due to symptom relapse in schizophrenia patients.[5] Timely intervention is crucial when a patient with a mental health disorder is at risk for an adverse event. Detecting changes in behavior as close to the time of onset as possible is vital so that a patient's psychiatric team can intervene to prevent a potential adverse event.

Though outpatient psychotherapy and psychiatric medication management can help control patients' symptoms, supplemental tools may be necessary to mitigate the risk of relapse or self-harm. With smartphones becoming more common, they provide an opportunity to monitor a patient's symptoms and establish a baseline for how a patient behaves. Smartphone data takes two primary forms: active data and passive data.[1] Active data requires patients' active input, such as answering self-report questionnaires about their symptoms on a given day. However, assessing one's own symptoms and overall well-being is challenging and subjective. Additionally, asking a patient to provide daily reports increases patient burden and patients may not be willing to provide this on a daily basis. Passive data, on the other hand, does not require active input from the patient. Examples of passive data collected from smartphone sensors are GPS information, such as total distance traveled on a given day, and social data, including the number of outgoing phone calls made each day. This type of data is collected daily without adding to patient burden and provides a more objective measure of how a patient is operating, as changes in mobility, activity, phone usage, or other behavioral patterns captured by smartphone sensors have been linked to mood and symptom changes in prior work.[6–8] Identifying changes in these behavioral patterns over time motivates the use of change point detection in mHealth applications.

Change point detection is the process of finding distributional changes in time-ordered data streams. These changes could be in mean, variance, correlation structure, or any other statistical properties of the underlying distributions. In mobile and behavioral health settings, such changes may reflect clinically meaningful shifts in behavior, such as reduced mobility, fewer outgoing calls, or changes in daily routines that can signal symptom worsening or relapse. Change point detection has also been used in other applications, including quality control and finance.[9–11] When there are multiple data streams that coincide, this is referred to as multivariate change point detection. As the number of data streams increases, more observations are required to have adequate power to detect change points.

Change point detection methods are often categorized as offline or online. Offline change point detection considers the entire data set retrospectively and evaluates candidate change points using observations both before and after the date in question. Online change point detection, on the other hand, treats the data as arriving sequentially and evaluates each potential change point using only the data observed before the date in question. Once a change is detected, the pre-change point days are removed and the monitoring process is restarted. In mHealth applications, where timely intervention is critical, the online setting is particularly relevant.

Several methods exist for multivariate change point detection. Hotelling's $T^2$ is an anomaly detection method developed in the context of multivariate quality control.[12] It is used to identify deviations in a data process from the expected mean assuming a constant covariance structure, most commonly in an offline setting. While this method is useful for detecting outliers, it is not designed or adequately powered to detect more subtle changes that persist over time.

To detect changes in time-ordered data that persist over time, cumulative sum (CUSUM) methods have been widely used. CUSUM was initially used for change point detection in univariate data in a quality control setting,[13] and was later extended for multivariate settings.[14] CUSUM methods are well-powered for multivariate change point detection in online settings; however, they can be sensitive to tuning parameters selected. Another drawback to CUSUM methods is that its power for detecting change points diminishes as the number of data streams increases.[15,16]

Sample divergence is a nonparametric method often used in offline multivariate change point detection.[17] This method is appealing when distributional assumptions cannot be made. However, since this method involves pairwise comparisons of observations, it can be computationally expensive.

This paper introduces a variance component score test for detecting changes in mean and/or variance in smartphone data. Hypothesis testing for variance components is popular in the context of mixed-effects modeling[18,19] and has also been used in genome-wide association studies (GWAS).[20] However, to our knowledge, it has not been used in the context of change point detection.

In this paper we introduce VC*, a method for multivariate change point detection. We evaluate the performance of VC* against existing methods in simulation studies. Lastly, we apply VC* to smartphone sensor data in a cohort of adolescents and young adults with affective instability.

## 2 Methods

### 2.1 Data and notation

We consider a single patient with $T$ days of follow-up and $p$ smartphone features. A change point occurs on day $k$, which is treated as the first post-change point day. On day $t \in \{1, 2, \ldots, T\}$, we have the following model:

$$\boldsymbol{Y}_t = \boldsymbol{\mu}_t + \mathbb{I}_{t \geq k}\boldsymbol{\beta} + \boldsymbol{\epsilon}_t, \quad \boldsymbol{\epsilon}_t \sim N_p\left(\boldsymbol{0}, \Sigma + \mathbb{I}_{t \geq k}\delta I_p\right)$$

where $\boldsymbol{Y}_t = [Y_{1t}, \ldots, Y_{pt}]^\top$ is the feature vector on day $t$, $\boldsymbol{\mu}_t = [\mu_{1t}, \ldots, \mu_{pt}]^\top$ is a vector of feature means, $\boldsymbol{\beta} = [\beta_1, \ldots, \beta_p]^\top$ is a vector containing shifts in each feature mean post-change point, $\Sigma$ is the pre-change point covariance matrix, $\delta$ is the shift in variance post-change point, and $\boldsymbol{\epsilon}_t = [\epsilon_{1t}, \ldots, \epsilon_{pt}]^\top$ is a vector of error terms. The vector of feature means can also be represented by $\boldsymbol{\mu}_t = \boldsymbol{\alpha} + \boldsymbol{\gamma}_t$, where $\boldsymbol{\alpha} = [\alpha_1, \ldots, \alpha_p]^\top$ is a vector of intercept coefficients and $\boldsymbol{\gamma}_t = [\gamma_{1t}, \ldots, \gamma_{pt}]^\top$ are weekly periodic effects such that $\sum_{t=1}^{7} \boldsymbol{\gamma}_t = \boldsymbol{0}$ and $\boldsymbol{\gamma}_t = \boldsymbol{\gamma}_{t+7}$ for all $t$. For ease of notation, we assume $\boldsymbol{\mu}_t = \boldsymbol{\mu}$ for all $t$; in other words, we assume that the feature mean vector is the same for each day of the week or that day-of-week effects have been subtracted out. Lastly, we let $db$ be the number of days back we search over for a change point. For example, $db = 7$ corresponds to searching for a change point over the last week.

We wish to test the null hypothesis $H_0: \{\boldsymbol{\beta} = \mathbf{0}\} \cap \{\delta = 0\}$, which corresponds to no change in the mean or variance.

## 2.2 Competing methods

### 2.2.1 Hotelling's $T^2$ max test

Hotelling's $T^2$ is an anomaly detection method derived under a multivariate normal distribution and based on a likelihood ratio test, originally designed to identify single outlier days in offline settings. We adapt it to our online setting by constructing a max test based on the Hotelling's $T^2$ test statistic. Let $\widehat{\boldsymbol{\mu}}$ be the overall sample mean of the features across all $T$ days, and let $\widehat{\Sigma}$ be the sample covariance matrix over all $T$ days. For a Hotelling's $T^2$ max text, we reject $H_0$ for large values of

$$max_{T-db \leq t \leq T-1}(\boldsymbol{Y}_t - \widehat{\boldsymbol{\mu}})^\top \widehat{\Sigma}^{-1}(\boldsymbol{Y}_t - \widehat{\boldsymbol{\mu}})$$

### 2.2.2 Sample divergence

Sample divergence is a nonparametric method often used in offline change point detection that measures dissimilarity between the distributions of data segments on either side of a candidate change point day. For notational simplicity, we reparameterize around $k$ and let $\boldsymbol{X}_t = \boldsymbol{Y}_t$ for $1 \leq t < k$ and $\boldsymbol{Z}_{t-k+1} = \boldsymbol{Y}_t$ for $k \leq t \leq T$. Moreover, let $T_0 = k - 1$ denote the number of days pre-change point and $T_1 = T - k + 1$ denote the number of days post-change point. For sample divergence, we reject $H_0$ for large values of

$$\frac{2}{T_0 T_1} \sum_{i=1}^{T_0} \sum_{j=1}^{T_1} \|\boldsymbol{X}_i - \boldsymbol{Z}_j\|_1 - \frac{2}{T_0(T_0-1)} \sum_{1 \leq i < j \leq T_0} \|\boldsymbol{X}_i - \boldsymbol{X}_j\|_1 - \frac{2}{T_1(T_1-1)} \sum_{1 \leq i < j \leq T_1} \|\boldsymbol{Z}_i - \boldsymbol{Z}_j\|_1$$

Unlike the other competing methods in this paper, the sample divergence test statistic is not standardized. So, we adjust the above sample divergence test statistic by taking its associated empirical p-value. Specifically, for each candidate change point day $k \in \{T - db, T - db + 1, \ldots, T - 1\}$, we first compute the observed sample divergence test statistic. Next, we generate 1000 data sets under the null hypothesis (no change point) and compute the corresponding sample divergence test statistics for each candidate $k$ in each data set. We then compute the p-value for day $k$, defined as the proportion of null test statistics on day $k$ that are greater than or equal to the observed test statistic on day $k$. The adjusted sample divergence test statistic is the candidate day with the minimum p-value. We reject $H_0$ for small values of the adjusted sample divergence test statistic.

### 2.2.3 CUSUM

Multivariate CUSUM is an online method for change point detection that accumulates deviations from a baseline (pre-change point) mean. Let $a > 0$ be a tuning parameter that determines the scaling and resetting of the CUSUM statistic. Crosier (1988) examined the choice of this tuning parameter through simulation studies of average run length and found $a = \sqrt{p}$ to perform well; we follow this recommendation here. Let $\boldsymbol{S}_t = (\boldsymbol{S}_{t-1} + |\boldsymbol{Y}_t - \widehat{\boldsymbol{\mu}}|)\left(1 - \frac{a}{b_t}\right)$ if $b_t > a$ and $\boldsymbol{S}_t = 0$ otherwise, where $b_t = (\boldsymbol{S}_{t-1} + \boldsymbol{Y}_t - \widehat{\boldsymbol{\mu}})^\top \Sigma^{-1} (\boldsymbol{S}_{t-1} + \boldsymbol{Y}_t - \widehat{\boldsymbol{\mu}})$. Then we reject $H_0$ for large values of

$$max_{T-db \leq t \leq T-1} \boldsymbol{S}_t^\top \widehat{\Sigma}^{-1} \boldsymbol{S}_t$$

## 2.3 Variance component score test

Here we extend the variance component score test framework, traditionally used for testing random effects, to the multivariate change point detection setting. Assume $\boldsymbol{\beta} \sim N_p(\boldsymbol{0}, \tau \boldsymbol{I}_p)$ for some finite $\tau > 0$ where $\boldsymbol{I}_p$ is a $p \times p$ identity matrix. Then testing $H_0: \{\boldsymbol{\beta} = \boldsymbol{0}\} \cap \{\delta = 0\}$ versus $H_a: \{\boldsymbol{\beta} \neq \boldsymbol{0}\} \cup \{\delta \neq 0\}$ is equivalent to testing $H_0: \{\tau = 0\} \cap \{\delta = 0\}$ versus $H_a: \{\tau \neq 0\} \cup \{\delta \neq 0\}$. This assumption reduces the problem from testing $p + 1$ parameters, $\{\beta_1, \ldots, \beta_p, \delta\}$, to testing two parameters, $\{\tau, \delta\}$.

Note that $\boldsymbol{Y}_1, \ldots, \boldsymbol{Y}_{k-1}$ are independent and identically distributed $N_p(\boldsymbol{\mu}, \Sigma)$ and $\boldsymbol{Y}_k|\boldsymbol{\beta}, \ldots, \boldsymbol{Y}_T|\boldsymbol{\beta}$ are independent and identically distributed $N_p(\boldsymbol{\mu} + \boldsymbol{\beta}, \Sigma + \delta \boldsymbol{I}_p)$. Thus, the likelihood of $\boldsymbol{Y} = [\boldsymbol{Y}_1, \ldots, \boldsymbol{Y}_T]$ is given by

$$L(\tau, \delta \mid \boldsymbol{Y}) = \prod_{t=1}^{k-1} f(\boldsymbol{y}_t) \times \int \left(\prod_{t=k}^{T} f(\boldsymbol{y}_t \mid \boldsymbol{\beta})\right) f(\boldsymbol{\beta}) \, d\boldsymbol{\beta}$$

$$\propto |\Sigma + \delta \boldsymbol{I}_p|^{-\frac{T-k+1}{2}} \tau^{-\frac{p}{2}} \exp\left\{-\frac{1}{2}\sum_{t=k}^{T}(\boldsymbol{y}_t - \boldsymbol{\mu})^\top (\Sigma + \delta \boldsymbol{I}_p)^{-1}(\boldsymbol{y}_t - \boldsymbol{\mu})\right\} \times$$

$$\int \exp\left(-\frac{1}{2}\left[\boldsymbol{\beta}^\top\left\{\left(\frac{\Sigma + \delta \boldsymbol{I}_p}{T-k+1}\right)^{-1} + \frac{\boldsymbol{I}_p}{\tau}\right\}\boldsymbol{\beta} - 2\boldsymbol{\beta}^\top (\Sigma + \delta \boldsymbol{I}_p)^{-1}\sum_{t=k}^{T}(\boldsymbol{y}_t - \boldsymbol{\mu})\right]\right) d\boldsymbol{\beta}$$

(derivation provided in the Supplemental Materials Section S1). Let $\boldsymbol{U}_k(\tau, \delta)$ and $\boldsymbol{I}_k(\tau, \delta)$ be the corresponding score vector and information matrix, respectively. The variance component (VC) score statistic for day $k$ is then given by

$$Q_k = \boldsymbol{U}_k^\top(\tau, \delta) \boldsymbol{I}_k^{-1}(\tau, \delta) \boldsymbol{U}_k(\tau, \delta)$$

Since it is unknown whether there is a change point, and if so on which day, we compute $Q_k$ for $T - db \leq k \leq T - 1$, which ensures that there are at least two days of post-change point data available at the candidate change point $k$. So, we reject $H_0$ for large values of

$$max_{T-db \leq k \leq T-1} Q_k$$

## 2.4 Estimating $\mu$ and $\Sigma$

In a typical variance component score test, $\mu$ and $\Sigma$ are estimated assuming the null hypothesis is true, which implies using all days of follow-up:

$$\hat{\mu} = \frac{1}{T} \sum_{t=1}^{T} y_t$$

$$\hat{\Sigma} = \frac{1}{T-1} \sum_{t=1}^{T} (y_t - \hat{\mu})(y_t - \hat{\mu})^\top$$

However, if the null hypothesis is not true and there is indeed a change point on day $k$, our estimates for $\mu$ and $\Sigma$ would be biased given the distribution change. This is outlined in the following theorem (proof provided in Supplementary Material Section S2).

**Theorem 2.1** Given that there is a change point on day $k$, $\hat{\mu}$ and $\hat{\Sigma}$ are biased estimators of $\mu$ and $\Sigma$, respectively. Specifically, $Bias_\mu(\hat{\mu} \mid \beta) = \frac{T-k+1}{T} \beta$ and $Bias_\Sigma(\hat{\Sigma} \mid \beta) = \frac{T-k+1}{T} \delta I_p + \frac{(k-1)(T-k+1)}{T(T-1)} \beta\beta^\top$.

Note that when there is a change point in the mean only ($\tau > 0$ and $\delta = 0$), both $\hat{\mu}$ and $\hat{\Sigma}$ are biased estimators; when there is a change point in the variance only ($\tau = 0$ and $\delta > 0$), only $\hat{\Sigma}$ is biased. Given this bias under $H_a$, we propose alternative estimators for $\mu$ and $\Sigma$ that remove this bias with a small cost to variance. We propose using only pre-change point days to estimate $\mu$ and $\Sigma$:

$$\hat{\mu}^* = \frac{1}{k-1} \sum_{t=1}^{k-1} y_t$$

$$\hat{\Sigma}^* = \frac{1}{k-2} \sum_{t=1}^{k-1} (y_t - \hat{\mu}^*)(y_t - \hat{\mu}^*)^\top$$

As this method is motivated by detecting change points in smartphone data with potentially highly correlated features or with more features than days ($p > T$), we allow for a regularization of our unbiased estimator of $\Sigma$. Let $V = diag(\sigma_1^2, \ldots, \sigma_p^2)$ be a diagonal matrix where element $[l, l]$ is the variance of smartphone feature $l \in \{1, \ldots, p\}$. Let $\hat{V} = diag(s_1^2, \ldots, s_p^2)$ be the unbiased estimator of $V$ using pre-change point days. Let

$$\hat{R} = \begin{pmatrix} 1 & r_{1,2} & \cdots & r_{1,p} \\ r_{2,1} & 1 & \cdots & r_{2,p} \\ \vdots & \vdots & \ddots & \vdots \\ r_{p,1} & r_{p,2} & \cdots & 1 \end{pmatrix}$$

where $r_{i,j}$ is the sample correlation between features $i$ and $j$ computed using pre-change point days. Consider

$$\hat{\Sigma}^*(\phi) = \hat{V}^{\frac{1}{2}}[(1-\phi)\hat{R} + \phi I_p]\hat{V}^{\frac{1}{2}}$$

where $\phi \in [0, 1]$ is a regularization parameter. When $\phi = 0$, $\hat{\Sigma}^*(0) = \hat{V}^{\frac{1}{2}}\hat{R}\hat{V}^{\frac{1}{2}} = \hat{\Sigma}^*$; in other words, with no regularization this estimate is equivalent to the unbiased estimator of $\Sigma$ computed using pre-change point days. When $\phi = 1$, $\hat{\Sigma}^*(1) = \hat{V}^{\frac{1}{2}}I_p\hat{V}^{\frac{1}{2}} = \hat{V}$; with maximum regularization, this estimate reduces to the diagonal matrix of sample variances of the $p$ features based on pre-change point days. A simulation study exploring the relationship between correlation structure and $\phi$ is given in Supplementary Materials Section S3.

## 3 Simulations

### 3.1 Power simulations

We first compare the power of VC* to that of the following existing methods: Hotelling's $T^2$,[12] CUSUM,[14] and sample divergence.[17] We also compare our VC* method derived from the model given in Section 2.1 with analogous models for which there is just a change point in the mean and just a change point in the variance, denoted as VC* (mean only) and VC* (variance only), respectively.

For each simulation setting, we generate feature matrix $Y_{p \times T}$ such that $\mu = 0_{p \times 1}$ and $\Sigma = I_{p \times p}$. We consider $T \in \{30, 60, 90\}$ and $p = 50$; this choice was made so we have $T < p$, $T$ slightly larger than $p$, and $T > p$. We consider a change point in the mean only ($\tau > 0$ and $\delta = 0$) and a change point in the variance only ($\delta > 0$ and $\tau = 0$); we also consider a change point in both the mean and variance ($\tau > 0$ and $\delta > 0$), which we describe in Supplemental Materials Section S4. We also vary the proportion of the $p$ features that have a change point, denoted $\omega \in \{0.5, 1\}$.

Lastly, we vary the number of days post-change point, $T - k + 1 = k^* \in \{2, 3, \ldots, 7\}$. In the case where we just have a change in the mean or a change in the variance, for each combination of $T$, $\omega$, and type of change point, we choose $\tau$ or $\delta$ such that VC* has power of 80% at $k^* = 4$. For these simulations, we generate $Y_1, \ldots, Y_T$ independently and set $\phi = 1$ for all settings.

Simulation results are displayed in Figure 1. When there is a change in mean only ($\tau > 0$ and $\delta = 0$), VC* (mean only) has the highest power in most settings with VC* in close second, though sample divergence outperforms VC* when $T = 30$. When there is a change in variance only ($\delta > 0$ and $\tau = 0$), VC* (variance only) has the highest power in most settings with VC* in close second. These results show that VC* is well-powered compared to existing methods and is robust to different types of change points.

### 3.2 Comparing power for VC* and VC

By estimating $\mu$ and $\Sigma$ using only pre-change point days instead of all days observed, we are sacrificing lower estimator variance for reduced bias. To determine whether this trade-off is worth it when we have small $T$, we compare VC* to a traditional variance component score test, which we will refer to as VC, that uses all observed days to estimate $\mu$ and $\Sigma$.

We generate feature matrix $Y_{p \times T}$ such that $\mu = \mathbf{0}_{p \times 1}$. We consider $Y_{p \times T}$ such that $T \in \{6, 7, \ldots, 30\}$, $p = 15$, and $k^* = 4$. We also consider a change point in the mean only ($\tau > 0$ and $\delta = 0$) and a change point in the variance only ($\delta > 0$ and $\tau = 0$). We vary the level of exchangeable feature correlation, $\alpha \in \{0, 0.5, 0.8\}$. We choose $\tau$ or $\delta$ such that VC* has 80% power at $T = 20$ and $\alpha = 0$. For each setting, we set $\phi = 1$, which corresponds to full covariance matrix regularization such that the off-diagonals are set to 0.

Simulation results are displayed in Figure 2. Note that when the off-diagonals are set to 0 when the actual values are $\alpha = 0.5$ (middle column) and $\alpha = 0.8$ (right column), power for both VC* and VC are much lower than when $\alpha = 0$ (left column). When there is a change point in the mean and $\alpha = 0$ (no correlation across features), VC* and VC perform similarly. When there is a change point in the variance and $\alpha = 0$, VC outperforms VC* for small $T$, while power becomes more comparable starting at $T = 21$. For $\alpha > 0$, VC* consistently outperforms VC across all levels of feature correlation for both change in mean and change in variance.

## 4 Real data application

We apply change point detection to passively collected smartphone data in a cohort of patients with affective instability (description in Section 4.1). While the proposed method is intended for online use in real time, the analysis here is retrospective and the data are analyzed sequentially to mimic an online implementation. We perform change point detection at the daily level (pipeline described in Section 4.4) and explore whether change points in smartphone data are related to any clinical characteristics of the participants.

## 4.1 Participants and data acquisition

We apply our method to a cohort of 41 adolescents and young adults enrolled in a study investigating affective instability in youth.[21] This study was approved by the Institutional Review Board (IRB) of the University of Pennsylvania. All participants provided their written informed consent prior to participating. For minor participants, parents or guardians provided consent and minors provided assent. Descriptive statistics summarizing patient diagnoses are given in Xia et al. (2022).

Smartphone data were acquired through the Beiwe platform.[2] Participants downloaded the Beiwe application to their smartphones, and the application passively recorded Global Positioning System (GPS) and social data that were summarized into daily features (mobility features detailed in Supplement of Barnett and Onnela (2020)).[22]

## 4.2 Processing smartphone data

We consider segments of smartphone data that span at least 14 days and have no more than 3 consecutive days of missingness per segment. For patients with two such segments, we consider them separately. In each data segment, we impute missing days using the na.interp() function in R. From there, we use an adaptation of the processing pipeline previously used in Barnett et al. (2018) (see Supplemental Figure S3).[23] We first perform a rank-based inverse standard normal transformation on each feature. Next, we regress each feature on day of week. Since in many cases we do not have many observations for each day of the week, we impose an L2 penalty on the regression parameters. The regularization parameter for the L2 penalty was manually chosen to minimize the amount of autocorrelation in the first 7 lags of the time series of a few patients with many days of follow-up and no missingness. After fitting this ridge regression, we extract the corresponding residuals, which we use to perform change point detection. To generate null distributions for hypothesis testing, we permute the residuals.

## 4.3 Additional assessments

Upon enrollment, each participant completed a series of questionnaires about their symptoms. Details about each assessment are in Supplemental Table S1. Patients were also assessed for personality disorders.

## 4.4 Statistical analysis

We perform online change point detection on the smartphone data using VC*, Hotelling's $T^2$, CUSUM, and sample divergence. Here, we choose a run-in period of 7 pre-change point days and at least 2 potential post-change point days, where the change point day itself is counted as one of the post-change point days, so we begin testing at $T = 9$ days. If no change point is detected, the next day is added and we repeat the testing procedure. We focus on detecting change points that occur within a 7-day window, so we set $db = 7$ if we have at least 15 days of data. Otherwise, we set $db = T - 8$ so that we still have a 7-day run-in pre-change point period. For example, if we have 9 days, we set $db = 1$ so that we are only testing whether a change point occurs on day 8; if we have 10 days, we set $db = 2$ so that we are testing for a change point on day 8 or day 9; and so on. If a change point is detected, the pre-change point days are deleted and we begin the process again starting at the change point day. For all methods we use a conventional 0.05 significance threshold based on the empirical null distribution, which we apply sequentially over time.

When implementing VC* with more features than days ($p > T$), we let $\phi = 1$. Otherwise, we select a value of $\phi$ using the following steps: (1) set initial $\phi = 0.9$; (2) generate a null distribution of VC* test statistics with the current value of $\phi$ using $\hat{\mu}^*$ and $\hat{\Sigma}^*$; (3) find $\tau$ such that VC* has 80% power with given $\phi$; (4) compute VC* power (using $\tau$ from step 3) over a grid of $\phi \in \{0, 0.1, \ldots, 0.9\}$; (5) set $\phi$ that yields highest power; (6) repeat steps 2-5 five times, or until the same value of $\phi$ is selected at two consecutive iterations.

## 4.5 Results

A summary of observed smartphone data and change points detected by VC* and competing methods are displayed in Figure 3. A total of 43 segments of sensor data were analyzed across 41 patients. VC* detected 99 change points across 31 subjects, Hotelling's $T^2$ detected 31 change points across 17 patients, CUSUM detected 32 change points across 15 patients, and sample divergence detected 33 change points across 17 patients. VC* detected the most change points among the methods considered. While the ground truth is unknown in the real data, these results are consistent with the simulation results in Section 3.1, which showed higher power for VC* compared to competing methods. We used a conventional 0.05 significance threshold for all methods, which likely contributes to the overall number of change points.

To quantify the similarity of change point detection across the competing methods we compute a pairwise similarity metric, defined as the percentage of shared change points, between each pair of methods. A similarity score of 0 corresponds to no agreement between methods, and a similarity score of 1 corresponds to perfect agreement. We observed the highest agreement between Hotelling's $T^2$ and CUSUM (similarity = 0.15), followed by the second highest agreement between Hotelling's $T^2$ and sample divergence (similarity = 0.07). The lowest agreement was between VC* and CUSUM (similarity = 0.02).

We relate change points detected by VC* to baseline symptom scores by computing the Spearman correlation between rate of VC* change points in smartphone data and baseline clinical assessment scores. VC* smartphone change point rate for each patient is defined as the total number of change points detected by VC* divided by the total number of days with smartphone data (including imputed days). 95% confidence intervals (CIs) for the Spearman correlations were computed using 1000 bootstrapped samples. Figure 4 displays the Spearman correlations and corresponding 95% CIs.

VC* change point rate has a marginally significant positive correlation with Component 4 (habitual sleep efficiency) of the Pittsburgh Sleep Quality Index (PSQI) (Spearman correlation = 0.36, 95% CI: (0.02-0.65)).[24] This PSQI subscale assesses sleep efficiency, or the percentage of time spent asleep while in bed. A higher score for this subscale corresponds to being less sleep efficient. This result suggests that individuals with lower sleep efficiency tend to have more change points in their smartphone data.

VC* change point rate has a marginally significant negative correlation with the Experience Seeking portion of the Brief Sensation Seeking (BSS) Scale (Spearman correlation = -0.39, 95% CI: (-0.71, -0.03)).[25] A patient's score for the Experience Seeking portion of the BSS Scale is the average score for the following two questions: (1) "I would like to explore strange places," and (2) "I would like to take off on a trip with no pre-planned routes or timetables." Each item is scored on a scale from 1 ("strongly disagree") to 5 ("strongly agree"), and a higher score corresponds to greater levels of curiosity about new experiences and lifestyles. The significant negative correlation with VC* smartphone change point rate implies that individuals who are more experience seeking have less frequent changes in their smartphone sensor data. To emphasize the modest degree of association, these comparisons are no longer significant after correcting for multiple comparisons.

Next we examine the relationship between personality disorder diagnosis and rate of VC* change points in smartphone data. Of the 41 participants in this study, 16 had a diagnosed personality disorder (n=12 with borderline personality disorder, n=4 with personality disorder not otherwise specified) and 25 had either no personality disorder (n=7) or deferred diagnosis (n=18). Figure 5 displays box plots of the distribution of VC* change point rates by personality disorder diagnosis. There is no evidence of a difference in mean smartphone change point rate across personality disorder diagnosis groups (t-test p-value = 0.48). Note that the 18 patients with deferred diagnosis were not evaluated for a personality disorder, which may attenuate any true difference.

## 5 Discussion

In this paper we introduce VC*, a variance component score test for multivariate change point detection for a mean vector, covariance matrix, or both. While variance component score tests have been used in other research areas, this is the first time it has been used in the context of multivariate change point detection. VC* is a novel implementation of a classic score test, as it estimates mean and variance parameters using pre-change point data to avoid biased estimators. Simulations show that this tradeoff of higher variance for lower bias results in higher power for change point detection. Additional power simulations demonstrate that VC* is better powered for detecting change points than existing methods. VC* is also robust to different types of change points, as it is well powered for detecting change points in mean vectors, covariance matrices, or both.

The competing methods considered in this paper were developed for different settings. Sample divergence and Hotelling's $T^2$ are most commonly used in offline settings to find change points retrospectively in a complete data set, whereas procedures such as CUSUM were designed for sequential monitoring in an online setting. In this work, we implement online versions of all methods to enable direct comparisons with VC*.

Our simulation studies were conducted in an offline setting to allow for controlled power comparisons across methods. In contrast, the real-data analysis focuses on online change point detection motivated by mHealth applications. Our results, along with prior literature, suggest that smartphone data can be meaningfully connected to clinical phenotypes.[2–4,23] In the mHealth setting in high-risk psychiatric populations, timely detection of behavioral shifts is crucial. Online change point detection methods are therefore well-suited to this context, as they aim to identify changes as they occur rather than retrospectively. VC* is particularly useful in this context as it is designed to detect recent deviations from pre-change point data, enabling immediate action when these changes are detected.

Implementing online change point detection in a real-data setting presents several challenges. When the number of pre-change point days is small, estimation of the mean and covariance is difficult. This challenge is relevant in monitoring clinical populations, where early detection is important but limited data are available prior to a change. Though VC* uses unbiased estimators of mean and covariance, performance still depends on having enough data to estimate these parameters reliably. In high-dimensional settings with many features and few days, regularization can help stabilize the estimated correlation matrix; however, the regularized estimate may not reflect the true covariance structure of the data, and power remains limited.

Finally, the choice of significance level for hypothesis testing in this setting remains an open question. We used the conventional 0.05 threshold, but this may not be appropriate in the context of sequential testing. In particular, the relatively large number of change points detected by VC* in the real-data analysis likely reflects the use of a fixed threshold applied repeatedly over time. In our setting, hypothesis tests are conducted on overlapping sets of data, so the resulting test statistics will inherently be correlated. Future applications may adjust the significance level to account for the correlation between the repeated hypothesis tests and control the probability of a false alarm. This is especially important in clinical monitoring, as too many false positives could strain resources or lead to participants disengaging.

## Funding

This work was funded in part by the A.E. Foundation, the Penn-CHOP Lifespan Brain Institute, and the National Institutes of Health (R01-MH107703 and R01-MH113550).

## Conflict of Interest

The authors declared no potential conflicts of interest with respect to the research, authorship, and/or publication of this article.

## Data availability

Data used in this article are available.

## References

1. Torous J, Staples P, Onnela JP. Realizing the Potential of Mobile Mental Health: New Methods for New Data in Psychiatry. *Curr Psychiatry Rep* 2015; 17: 61.

2. Torous J, Kiang MV, Lorme J, et al. New Tools for New Research in Psychiatry: A Scalable and Customizable Platform to Empower Data Driven Smartphone Research. *JMIR Ment Health* 2016; 3: e5165.

3. Mohr DC, Zhang M, Schueller SM. Personal Sensing: Understanding Mental Health Using Ubiquitous Sensors and Machine Learning. *Annu Rev Clin Psychol* 2017; 13: 23–47.

4. Onnela JP, Rauch SL. Harnessing Smartphone-Based Digital Phenotyping to Enhance Behavioral and Mental Health. *Neuropsychopharmacology* 2016; 41: 1691–1696.

5. Torous J, Firth J, Mueller N, et al. Methodology and Reporting of Mobile Health and Smartphone Application Studies for Schizophrenia. *Harv Rev Psychiatry* 2017; 25: 146.

6. Rohani DA, Faurholt-Jepsen M, Kessing LV, et al. Correlations Between Objective Behavioral Features Collected From Mobile and Wearable Devices and Depressive Mood Symptoms in Patients With Affective Disorders: Systematic Review. *JMIR MHealth UHealth* 2018; 6: e165.

7. Pratap A, Atkins DC, Renn BN, et al. The accuracy of passive phone sensors in predicting daily mood. *Depress Anxiety* 2019; 36: 72–81.

8. Ringwald WR, King G, Vize CE, et al. Passive Smartphone Sensors for Detecting Psychopathology. *JAMA Netw Open* 2025; 8: e2519047.

9. Habibi R. Bayesian Online Change Point Detection in Finance. *Financ Internet Q* 2022; 17: 27–33.

10. Pepelyshev A, Polunchenko AS. Real-time financial surveillance via quickest change-point detection methods. Epub ahead of print 2015. DOI: 10.48550/arXiv.1509.01570.

11. Kim K, Park JH, Lee M, et al. Unsupervised Change Point Detection and Trend Prediction for Financial Time-Series Using a New CUSUM-Based Approach. *IEEE Access* 2022; 10: 34690–34705.

12. Hotelling H. Multivariate Quality Control. *Tech Stat Anal*, https://cir.nii.ac.jp/crid/1572824500399407104 (1947, accessed 20 April 2024).

13. Page ES. Continuous Inspection Schemes. *Biometrika* 1954; 41: 100–115.

14. Crosier RB. Multivariate Generalizations of Cumulative Sum Quality-Control Schemes. *Technometrics* 1988; 30: 291–303.

15. Malinas R, Song D, Robinson BD, et al. High-Dimensional Sequential Change Detection. Epub ahead of print 2025. DOI: 10.48550/arXiv.2502.05377.

16. Liu B, Zhang X, Liu Y. High Dimensional Change Point Inference: Recent Developments and Extensions. *J Multivar Anal* 2022; 188: 104833.

17. Matteson DS, James NA. A Nonparametric Approach for Multiple Change Point Analysis of Multivariate Data. *J Am Stat Assoc* 2014; 109: 334–345.

18. Lin X. Variance component testing in generalised linear models with random effects. *Biometrika* 1997; 84: 309–326.

19. Verbeke G, Molenberghs G. The Use of Score Tests for Inference on Variance Components. *Biometrics* 2003; 59: 254–262.

20. Tzeng JY, Zhang D. Haplotype-Based Association Analysis via Variance-Components Score Test. *Am J Hum Genet* 2007; 81: 927–938.

21. Xia CH, Barnett I, Tapera TM, et al. Mobile footprinting: linking individual distinctiveness in mobility patterns to mood, sleep, and brain functional connectivity. *Neuropsychopharmacology* 2022; 47: 1662–1671.

22. Barnett I, Onnela JP. Inferring mobility measures from GPS traces with missing data. *Biostatistics* 2020; 21: e98–e112.

23. Barnett I, Torous J, Staples P, et al. Relapse prediction in schizophrenia through digital phenotyping: a pilot study. *Neuropsychopharmacol Off Publ Am Coll Neuropsychopharmacol* 2018; 43: 1660–1666.

24. Buysse DJ, Reynolds CF, Monk TH, et al. The Pittsburgh Sleep Quality Index: a new instrument for psychiatric practice and research. *Psychiatry Res* 1989; 28: 193–213.

25. Hoyle RH, Stephenson MT, Palmgreen P, et al. Reliability and validity of a brief measure of sensation seeking. *Personal Individ Differ* 2002; 32: 401–414.

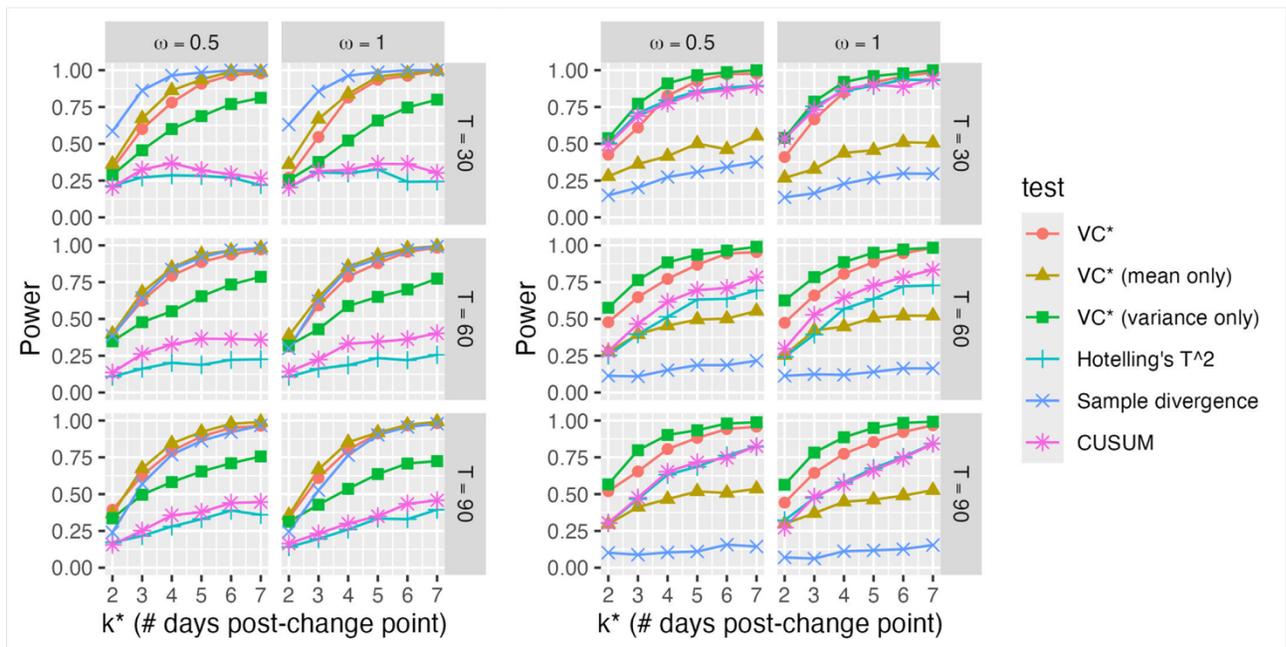

**Figure 1** Power simulations for (a) change in mean and (b) change in variance. Power curves are shown over the number of days post-change point $k^*$, stratified by total time points $T$ and proportion of affected features $\omega$. Each simulated data set has $p = 50$ features.

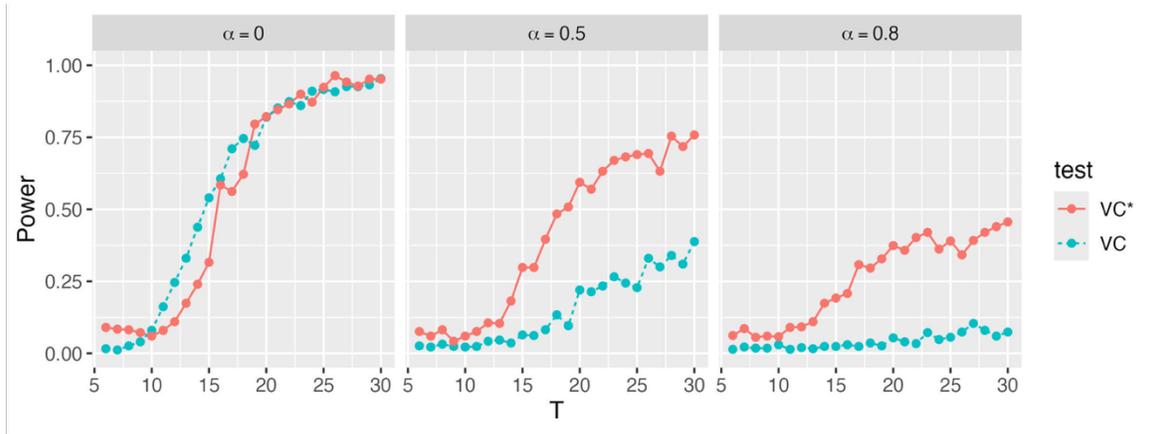

(a) Change in mean

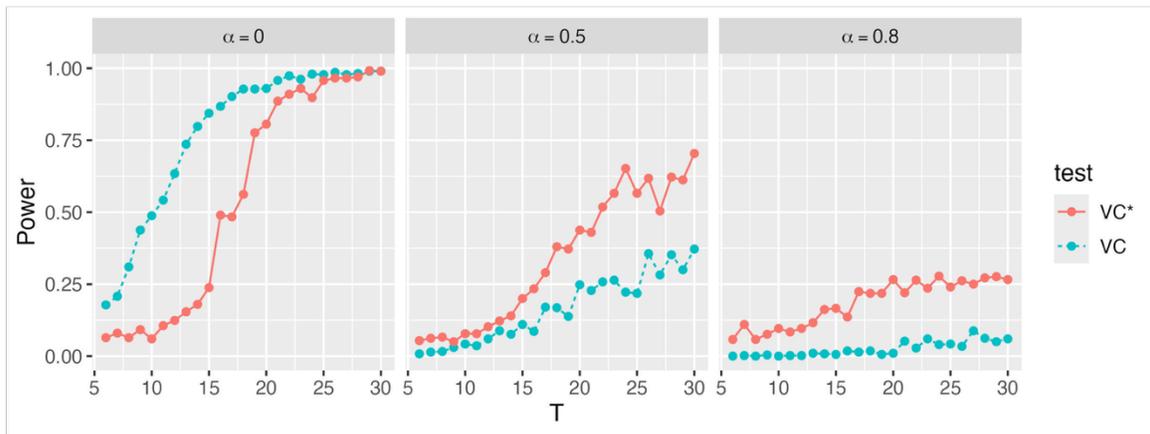

(b) Change in variance

**Figure 2** Power simulations comparing VC* and VC for small $T$ when there is a (a) change in mean and (b) change in variance. Data are generated with $p = 15$ features and $k^* = 4$. We vary the level of exchangeable feature correlation $\alpha \in \{0, 0.5, 0.8\}$. We choose the change point effect size such that VC* has 80% power at $T = 20$ and $\alpha = 0$.

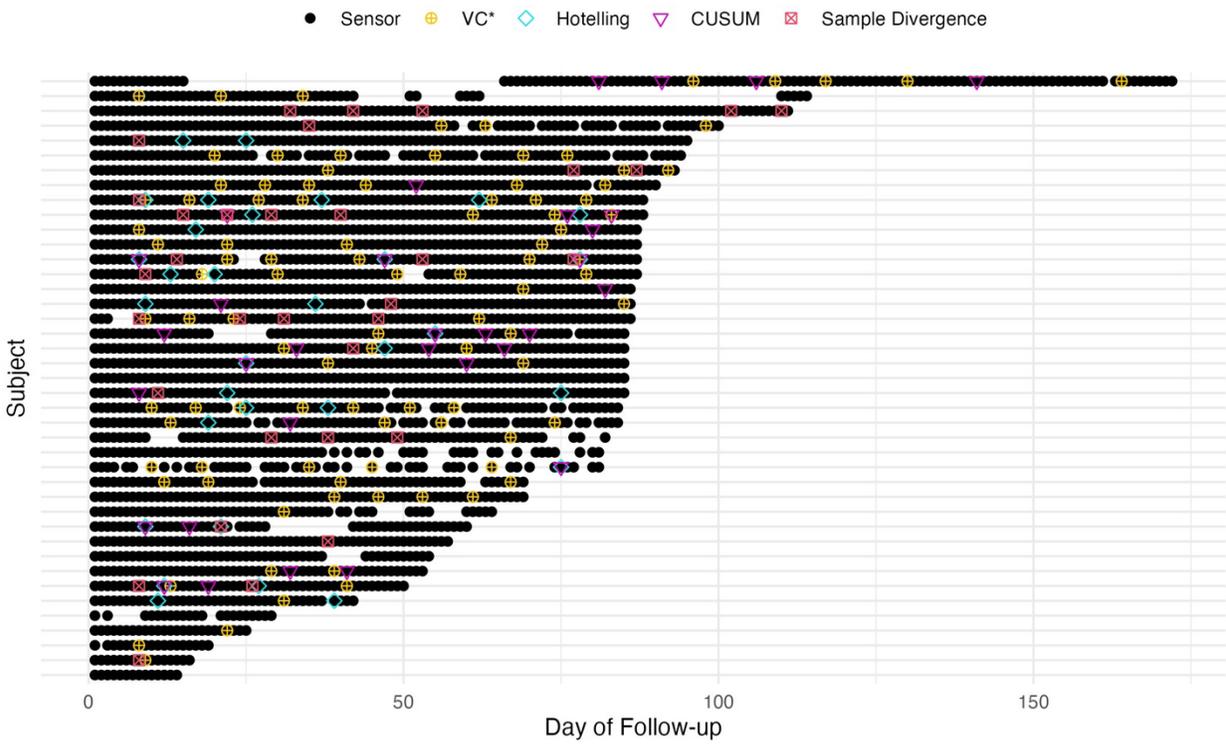

**Figure 3** Summary of smartphone sensor data and change points detected across subjects. Black dots represent days with passively collected smartphone data. Change points are marked with distinct shapes: a yellow circle with a cross for VC*, a blue diamond for Hotelling's $T^2$, a purple triangle for CUSUM, and a red square with an X for sample divergence.

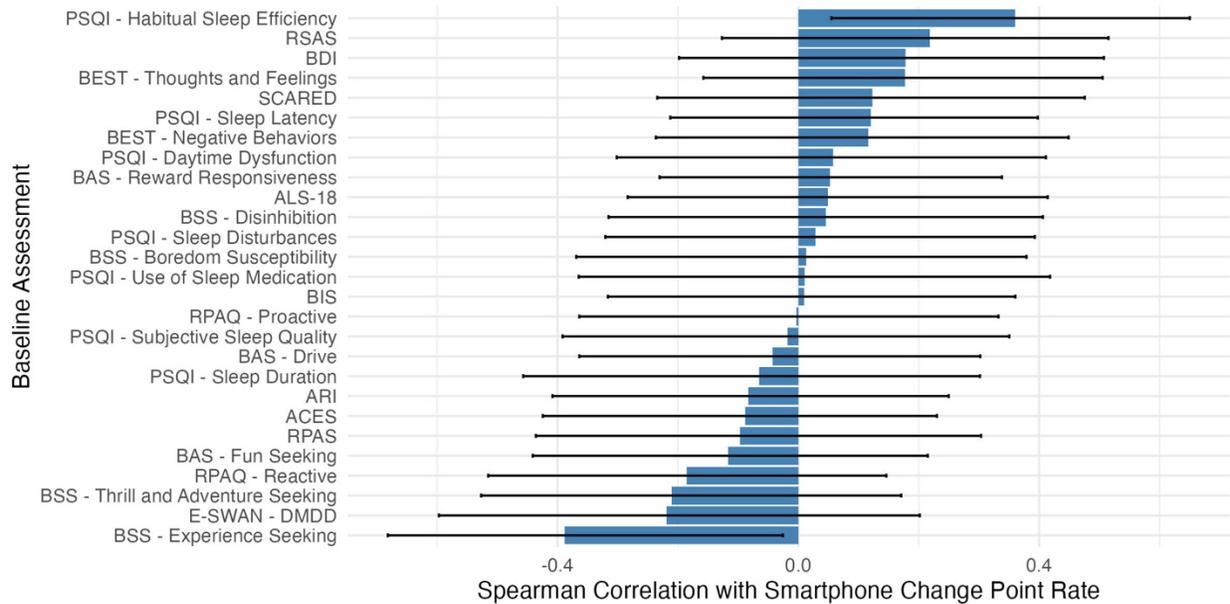

**Figure 4** Spearman correlation between rate of VC* change points and baseline clinical assessment scores. VC* smartphone change point rate for each patient is defined as the total number of change points detected by VC* divided by the total number of days with smartphone data (including imputed days). 95% confidence intervals were computed using bootstrapping.

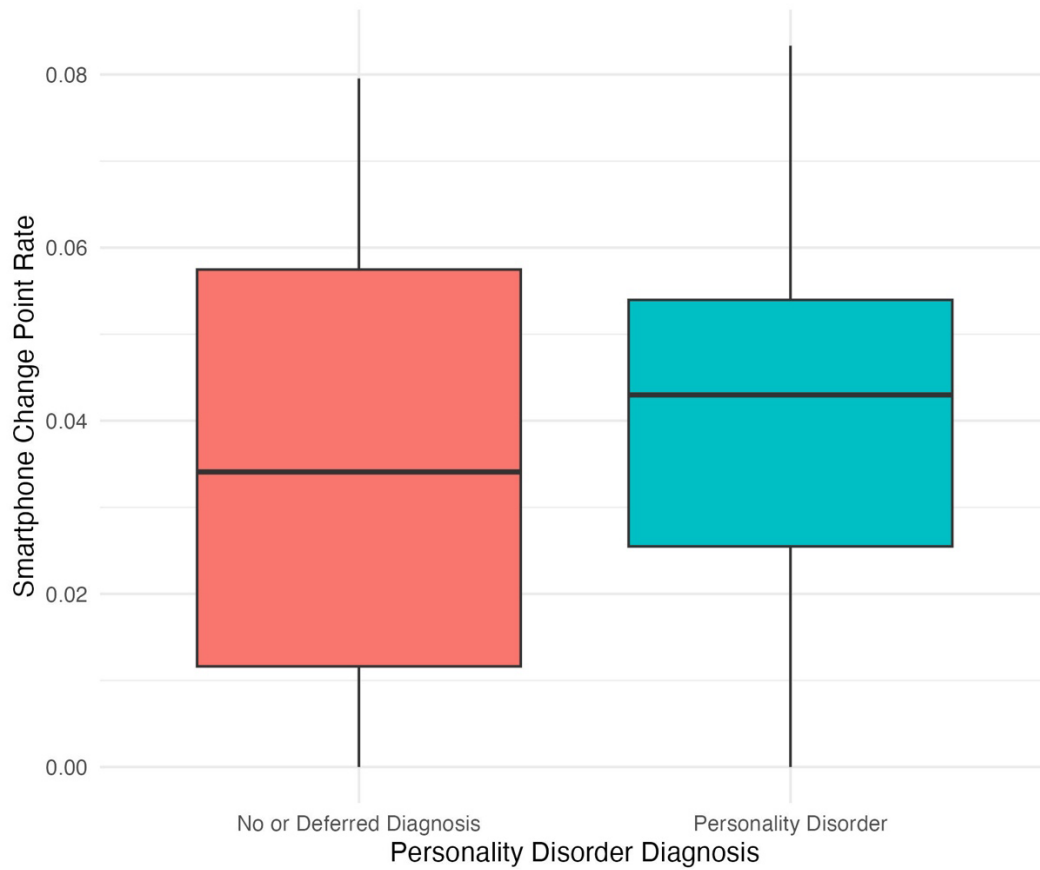

**Figure 5** Box plots showing the distribution of VC* change point rate by personality disorder diagnosis. VC* smartphone change point rate for each patient is defined as the total number of change points detected by VC* divided by the total number of days with smartphone data (including imputed days). Of the 41 patients included in this study, 16 had a diagnosed personality disorder (12 with borderline personality disorder, 4 with personality disorder not otherwise specified) and 25 had either no diagnosis (n=7) or deferred diagnosis (n=18).